\documentclass[final,3p,times,twocolumn]{elsarticle}

\usepackage{graphicx}

\usepackage{amsmath,amssymb}

\biboptions{comma,sort&compress}

\journal{Nuc. Phys. (Proc. Suppl.)}

\newcommand{\Pis}{\Pi_{\text{(S)}}}
\newcommand{\borel}{\hat{\mathcal{B}}}
\newcommand{\laplace}{\mathcal{L}}

\begin{document}

\begin{frontmatter}



\title{A Laplace Sum-Rules Analysis of Heavy Pseudoscalar ($J^{PC}=0^{-+}$) Hybrids}


\author[label2]{D.\ Harnett\corref{cor1}}\ead{derek.harnett@ufv.ca}
\author[label2]{R.\ Berg}
\author[label3]{R.T.\ Kleiv}
\author[label3]{T.G.\ Steele}

\address[label2]{Department of Physics, University of the Fraser Valley, Abbotsford, BC, V2S 7M8, Canada}
\address[label3]{Department of Physics and Engineering Physics, University of Saskatchewan, Saskatoon, SK, S7N 5E2, Canada}

\cortext[cor1]{Speaker}

\begin{abstract}
We use QCD sum-rules to predict ground state masses for pseudoscalar 
($J^{PC}=0^{-+}$) charmonium and bottomonium hybrids.  We find that the inclusion
of a six-dimensional gluon condensate contribution is needed to stabilize
the analyses. For the charmonium hybrid, we find a mass of $(3.82\pm0.13)$ GeV;
for the bottomonium hybrid, we find a mass of $(10.64\pm0.19)$ GeV.  We comment
on possible phenomenological implications concerning the $Y(3940)$.
\end{abstract}

\end{frontmatter}
Over the past decade, more than a dozen new charmonium-like states, 
the XYZ resonances, 
have been discovered (see~\cite{Eidelman:2012vu} for a recent review).
Few of these states can be easily accommodated with a conventional charmonium meson 
interpretation~\cite{Barnes:1995hc}.  There are discrepancies between observations
and calculations regarding masses and widths; certain $J^{PC}$ sectors seem to be 
overpopulated---the vector ($1^{--}$) states, in particular;
and electrically charged resonances
have perhaps been seen.  Naturally, there has been considerable speculation
that some of these new states may lie outside of the constituent quark model.

Motivated by these findings,
we use QCD sum-rules to predict ground state masses
of both pseudoscalar ($0^{-+}$) charmonium and bottomonium hybrids.
The first application of sum-rules to heavy hybrids was done 
in~\cite{Govaerts:1985fx,Govaerts:1984hc,Govaerts:1986pp}.
Therein, a variety of $J^{PC}$ quantum numbers were investigated;
however, in some cases (such as $1^{--}$ and $0^{-+}$), the sum-rules
were unstable and the resulting mass estimates were deemed unreliable.
Recently, some of these sum-rules have been re-analysed~\cite{Qiao:2010zh,Berg:2012gd,Harnett:2012gs}.
In the vector analysis of~\cite{Qiao:2010zh}, it was seen that 
the inclusion of a six-dimensional gluon condensate contribution 
stabilized the sum-rules and led to a reliable ground state mass prediction.
Therefore, in an effort to update the original pseudoscalar work in a similar fashion,
we also include a six-dimensional gluon condensate contribution.

As in~\cite{Govaerts:1985fx}, we define a current
\begin{equation}\label{current}
  j_\mu=\frac{g}{2}\bar Q\gamma^\nu\lambda^a\tilde G^a_{\mu\nu}Q,\
  \tilde G^a_{\mu\nu}=\frac{1}{2}\epsilon_{\mu\nu\alpha\beta}G^{a\,\alpha\beta}
\end{equation}
where $Q$ represents a heavy quark (charm or bottom) operator,
and we define a corresponding two-point function
\begin{align}
  \Pi_{\mu\nu}(q)&=i\!\int\! d^{\,4}\!x \,e^{i q\cdot x}
  \langle0\vert\, T j_\mu(x)j_\nu(0)\,|0\rangle\\
  \label{basic_corr}
  &=\Bigg( \frac{q_\mu q_\nu}{q^2}-g_{\mu\nu} \Bigg) \Pi_{\text{(V)}}\Big(q^2\Big)+\frac{q_\mu q_\nu}{q^2}\Pis\Big(q^2\Big).
\end{align}
(The S stands for scalar; the V for vector.)
Then, the longitudinal projection of $\Pi_{\mu\nu}$
\begin{equation}
  \Pis\Big(q^2\Big)=\frac{q^\mu q^\nu}{q^2}\Pi_{\mu\nu}(q)
\end{equation}
serves as a probe of heavy, pseudoscalar hybrids.

\begin{figure}[ht]\centering
  \includegraphics[scale=0.5]{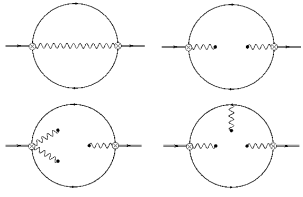}
  \caption{\label{feynman_diagrams} Feynman diagrams included in the OPE
    of $\Pis$.}
\end{figure}

We compute $\Pis$ using the operator product expansion (OPE) with contributions from 
perturbation theory, the four-dimensional gluon condensate 
\begin{equation}
\begin{split}
  \langle\alpha G^2\rangle&=\langle\alpha G^{a}_{\mu\nu} G^{a\,\mu\nu}\rangle\\
                          &=\left(7.5\pm 2.0\right)\times 10^{-2}\,{\rm GeV^4}
\end{split}
\label{GG_value}
\end{equation} 
as well as the six-dimensional gluon condensate 
\begin{equation}
\begin{split}
  \langle g^3 G^3\rangle&=\langle g^3 f^{abc}G^{a}_{\mu\nu}G^{b\,\nu}_{\ \ \rho}G^{c\,\rho\mu}\rangle\\
                       &=\left(8.2\pm 1.0\right){\rm GeV^2}\langle \alpha G^2\rangle,
\end{split}
\label{GGG_value}
\end{equation}
and we write
\begin{equation}
  \Pis^{\text{QCD}}\Big(q^2\Big)=\Pis^{\text{pert}}\Big(q^2\Big)+\Pis^{\text{GG}}\Big(q^2\Big)+\Pis^{\text{GGG}}\Big(q^2\Big).
\label{ope}
\end{equation}
The numerical values in~(\ref{GG_value}) and~(\ref{GGG_value}) are extracted from heavy 
quark systems~\cite{Narison:2010cg}.
At leading-order in $\alpha$, the relevant Feynman diagrams are shown in 
Figure~\ref{feynman_diagrams}\footnote{All Feynman diagrams are drawn using JaxoDraw\cite{Binosi:2003yf}.}.
Divergences are handled using dimensional regularization in $D=4+2\epsilon$ dimensions
followed by $\overline{\text{MS}}$-renormalization.
The Wilson coefficients of both gluon condensate terms are calculated using fixed-point
gauge methods (see~\cite{Elias:1987ac,Bagan:1992tg}, for example).
We use the program TARCER~\cite{Mertig:1998vk} to apply the two-loop integral recurrence relations
of~\cite{Tarasov:1996br,Tarasov:1997kx} which significantly reduces the number of integral formulae needed
to obtain exact results.  
The required two-loop integrals are in~\cite{Broadhurst:1993mw}; 
the required one-loop integrals are in~\cite{Davydychev:1991}.
Expanding in $\epsilon$ and omitting polynomials in $q^2$ as they ultimately provide no contribution to the sum-rules, we find
\begin{multline}
  \Pis^{{\rm pert}}\Big(q^2\Big)=\frac{\alpha m^6}{\pi^3}
  \Biggl[
  \frac{1}{30}(z-1)\left(4z^2-21z+10 \right)\\
  \times\phantom{}_3F_2\!\left(1,1,1;3/2, 3;z\right)\\
  +\frac{1}{270}z\left(8z^3+8z^2+29z-10  \right)\\
   \times\phantom{}_3F_2\!\left(1, 1, 2; 5/2, 4;z\right)
  \Biggr],
\label{Pi_pert}
\end{multline}
\begin{equation}
  \Pis^{\text{GG}}\Big(q^2\Big)=\frac{m^2}{18\pi} z(2z+1)\,
  \phantom{}_2F_1\!\left(1, 1; 5/2;z\right) \langle\alpha G^2\rangle,
\label{Pi_GG}
\end{equation}
and
\begin{multline}
  \Pis^{\text{GGG}}\Big(q^2\Big)=\frac{1}{384\pi^2 (z-1)^2}
  \Bigg[
    (2z^2-2z+1)\\
    \times\phantom{}_2F_1\!\left(1, 1; 5/2;z\right)\\
    +(10z^2-20z+7)
  \Bigg] \langle g^3G^3\rangle
\label{Pi_GGG}
\end{multline}
where $z=\frac{q^2}{4m^2}$ and where $\phantom{ }_p F_q(\cdots;\cdots;z)$ are
generalized hypergeometric functions (see~\cite{Bateman:1953}, for example).
For $q=p-1$, the function $\phantom{ }_p F_q(\cdots;\cdots;z)$ has a branch cut
discontinuity in the complex $z$-plane extending from $z=1$ to $z\rightarrow\infty$;
as such, it is readily seen from~(\ref{Pi_pert})--(\ref{Pi_GGG}) that $\Pis^{\text{QCD}}$ has a branch cut
extending from $q^2=4m^2$ to $q^2\rightarrow\infty$ as expected.

Due to its analytic structure and asymptotic behaviour, $\Pis$ satisfies the following
dispersion relation:
\begin{multline}
  \Pis\left(Q^2\right)=
  \frac{Q^8}{\pi}\!\int\limits_{4m^2}^\infty\!\frac{\mathrm{Im}\Pis(t)}{t^4\left(t+Q^2\right)}\,dt
  +\Pis(0)\\
  +Q^2\Pis'(0)+\frac{1}{2}Q^4\Pis''(0)+\frac{1}{6}Q^6\Pis'''(0)
\label{disp_rel}
\end{multline}
where $Q^2 =-q^2$.
To eliminate unwanted polynomials in $q^2$ and to suppress the high-energy (continuum) contributions 
from $\mathrm{Im}\Pis$ to the integral on the right-hand side of~(\ref{disp_rel}), we apply the Borel 
transform
\begin{equation}
  \borel\equiv\!\lim_{\stackrel{N,~Q^2\rightarrow \infty}{N/Q^2\equiv \tau}}
  \frac{\left(-Q^2\right)^N}{\Gamma(N)}\left(\frac{d}{dQ^2}\right)^N
\label{borel}
\end{equation}  
to obtain
\begin{equation}
  \frac{1}{\tau}\borel\,\Big[\Big(-Q^2\Big)^k\Pis\Big(Q^2\Big)\Big]
  =\frac{1}{\pi}\!\int_{4m^2}^{\infty}\!t^k e^{-t\tau}\mathrm{Im}\Pis(t)\,dt.
\label{midway}
\end{equation}
On the left-hand side of~(\ref{midway}), we use~(\ref{ope})--(\ref{Pi_GGG}) to approximate $\Pis$;
on the right-hand side, we employ a single narrow resonance plus continuum model
\begin{equation}
  \mathrm{Im}\Pis(t)=f^2\delta\left(t-M^2\right)
  +\theta\left(t-s_0\right)\mathrm{Im}\Pis^{\text{QCD}}(t)
\end{equation}
where $s_0$ is the continuum threshold parameter and $f$ represents a hadronic coupling.
Moving the threshold contribution from the right- to the left-hand side of~(\ref{midway}),
we arrive at the Laplace sum-rules~\cite{Shifman:1978bx,Shifman:1978by}
\begin{equation}
\begin{split}
  \laplace_k(\tau,\,s_0)\equiv&\frac{1}{\tau}\borel\,\Big[\Big(-Q^2\Big)^k\Pis^{\text{QCD}}\Big(Q^2\Big)\Big]\\
  &\qquad-\frac{1}{\pi}\!\int_{s_0}^{\infty}\!t^k e^{-t\tau}\mathrm{Im}\Pis^{\text{QCD}}(t)\,dt\\
  =&f^{2}M^{2k}e^{-M^2 \tau}.
\label{laplace_initial}
\end{split}
\end{equation}
By exploiting a well-known relationship between the Borel transform and the inverse Laplace
transform~\cite{Bertlmann:1984ih}, we can simplify the remaining Borel transformed term 
in~(\ref{laplace_initial}) (see~\cite{Harnett:2000xz}, for example).
Doing so gives
\begin{multline}
  \laplace_0\left(\tau,s_0\right)=
    \frac{4m^2}{\pi}\!\int_1^{s_0/4m^2}\!e^{-4m^2\tau x}
     \left[{\rm Im}\Pis^{\rm pert}\left(4m^2 x\right)\right.\\
    \left.+{\rm Im}\Pis^{\rm GG}\left(4m^2 x\right)\right]\,dx\\
    +\lim_{\eta\to 0^+}
     \left[\frac{4m^2}{\pi}\!\int_{1+\eta}^{s_0/4m^2}\!
     e^{-4m^2\tau x}\,{\rm Im}\Pis^{\rm GGG}(4m^2 x)\,dx\right.\\ 
    \left.-\frac{4m^2\langle g^3G^3\rangle}{128\pi^2\sqrt{\eta}}e^{-4m^2\tau}\right]
\label{L_0}
\end{multline}
and
\begin{equation}
\laplace_1\left(\tau,s_0\right)=-\frac{\partial}{\partial\tau}\laplace_0\left(\tau,s_0\right)
\label{L_1}
\end{equation}
where, for $z>1$,
\begin{multline}
  {\rm Im}\Pis^{\rm pert}\Big(q^2\Big)=\frac{\alpha m^6}{120\pi^2z^2}
  \Bigg[\sqrt{z-1} \sqrt{z}\\
    \times\left(30-115 z+166 z^2+8 z^3+16 z^4\right)\Biggr.\\
  -15 \left(-2+9 z-16 z^2+16 z^3\right) \log\left(\!\sqrt{z-1}+\sqrt{z}\,\right)
  \Biggr],
\label{Im_Pi_pert}
\end{multline}
\begin{equation}
  {\rm Im}\Pis^{\rm GG}\Big(q^2\Big)
  =\frac{m^2}{12}(2z+1)\frac{\sqrt{z-1}}{\sqrt{z}}\langle\alpha G^2\rangle,
  \label{Im_Pi_GG}
\end{equation}
and
\begin{multline}
  {\rm Im}\Pis^{\rm GGG}\Big(q^2\Big)=\frac{1}{256\pi z(z-1)^2}\frac{\sqrt{z-1}}{\sqrt{z}}\\
  \times(2z^2-2z+1)\langle g^3G^3\rangle.
\label{Im_Pi_GGG}
\end{multline}
 
We note that the portion of~(\ref{Pi_GGG}) that has a double pole at $z=1$ ultimately 
contributes to both~(\ref{L_0}) and~(\ref{L_1}); 
hence, $\mathrm{Im}\Pis^{\text{GGG}}$ alone is insufficient to 
formulate the sum-rules.  A full calculation of $\Pis^{\text{GGG}}$ is required.

In~(\ref{L_0}) and~(\ref{L_1}), the parameters $m$ and $\alpha$ are $\overline{\text{MS}}$-scheme
running quantities evaluated at a scale $\mu$.
For charm quarks, 
\begin{gather}
  \alpha(\mu)=\frac{\alpha\left(M_\tau\right)}{1+\frac{25\alpha\left(M_\tau\right)}{12\pi}
    \log{\left(\frac{\mu^2}{M_\tau^2}\right)}},\ 
    \alpha\left(M_\tau\right)=0.33\label{alphac}\\
  m_c(\mu)=\overline m_c\left(\frac{\alpha(\mu)}{\alpha\left(\overline m_c\right)}\right)^{12/25},\ 
  \overline m_c=\left(1.28\pm 0.02\right)\,{\rm GeV}.\label{massc}
\end{gather}
For bottom quarks
\begin{gather}
  \alpha(\mu)=\frac{\alpha\left(M_Z\right)}{1+\frac{23\alpha\left(M_Z\right)}{12\pi}
    \log{\left(\frac{\mu^2}{M_Z^2}\right)}},\ 
    \alpha\left(M_Z\right)=0.118\\
  m_b(\mu)=\overline m_b\left(\frac{\alpha(\mu)}{\alpha\left(\overline m_b\right)}\right)^{12/23},\ 
  \overline m_b=\left(4.17\pm 0.02\right)\,{\rm GeV}.\label{massb}
\end{gather}
In~(\ref{alphac})--(\ref{massb}), 
the quark mass parameters are from~\cite{Chetyrkin:2009fv,Narison:2011rn,Narison:2010cg,Kuhn:2007vp},
the $\tau$ and $Z$ masses are from~\cite{pdg},
and $\alpha(M_{\tau})$ and $\alpha(M_Z)$ are from~\cite{Bethke:2009jm}.
Furthermore, renormalization-group improvement sets $\mu=\frac{1}{\sqrt{\tau}}$~\cite{Narison:1981ts}.   

\begin{figure}[hbt]
\centering
\includegraphics[scale=0.8]{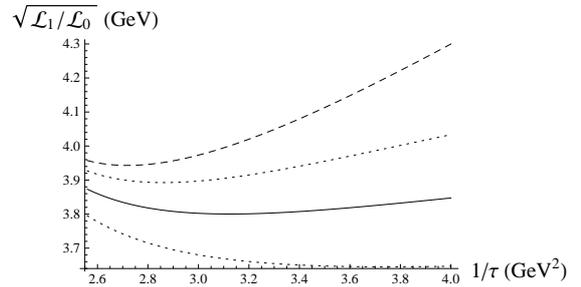}
\caption{The ratio ${\cal L}_1^{\rm QCD}\left(\tau,s_0\right)/{\cal L}_0^{\rm QCD}\left(\tau,s_0\right)$
for hybrid charmonium is shown  as a function of the Borel scale $1/\tau$ for
the optimized value $s_0=23\,{\rm GeV^2}$ (solid curve).  
For comparison the ratio is also shown for $s_0=28\,{\rm GeV^2}$ (upper dotted curve) and $s_0=19\,{\rm GeV^2}$ (lower dotted curve). 
The uppermost dashed curve represents the $s_0\to\infty$ limit corresponding to the bound $M<3.96\,{\rm GeV}$.
Central values of the QCD parameters have been used.}
\label{charm_opt}
\end{figure}

\begin{figure}[hbt]
\centering
\includegraphics[scale=0.8]{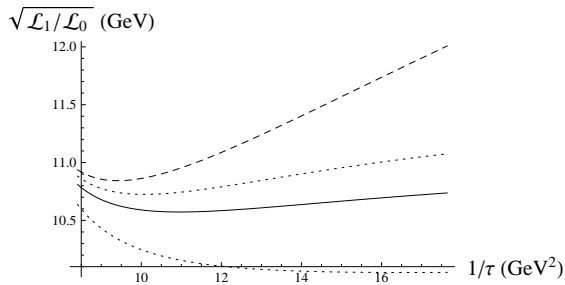}
\caption{The ratio ${\cal L}_1^{\rm QCD}\left(\tau,s_0\right)/{\cal L}_0^{\rm QCD}\left(\tau,s_0\right)$
for hybrid bottomonium is shown  as a function of the Borel scale $1/\tau$ for
the optimized value $s_0=140\,{\rm GeV^2}$ (solid curve).  
For comparison the ratio is also shown for $s_0=155\,{\rm GeV^2}$ (upper dotted curve) and $s_0=116\,{\rm GeV^2}$ (lower dotted curve). 
The uppermost dashed curve represents the $s_0\to\infty$ limit corresponding to the bound $M<10.84\,{\rm GeV}$.
Central values of the QCD parameters have been used.
}
\label{bottom_opt}
\end{figure}

From~(\ref{laplace_initial}), it follows that
\begin{equation}\label{sqrt_ratio}
  \sqrt{\frac{\laplace_1(\tau,\,s_0)}{\laplace_0(\tau,\,s_0)}}=M^2.
\end{equation}
The range of acceptable $\tau$-values is determined by demanding that, on the
left-hand side of~(\ref{sqrt_ratio}), the continuum contributes less than 30\%
and the condensates contribute less than 10\%.  Then, $s_0$ and $M$ are chosen 
to optimize~(\ref{sqrt_ratio}) over this range.  See~\cite{Berg:2012gd}
for further details.
In the charmonium hybrid case, we find $s_0=23\ \text{GeV}^2$ and 
$M=(3.82\pm0.13)\ \text{GeV}$; in the bottomonium hybrid case, we find
$s_0=140\ \text{GeV}^2$ and $M=(10.64\pm0.19)\ \text{GeV}$.
The uncertainties in these mass predictions are dominated by the 
uncertainties in the quark mass parameters~(\ref{massc}) and~(\ref{massb}) and
the uncertainties in the six-dimensional gluon condensate~(\ref{GGG_value}).
Plots of the left-hand side of~(\ref{sqrt_ratio}) for various values of $s_0$
are shown in Figures~\ref{charm_opt} and~\ref{bottom_opt} for charmonium and 
bottomonium hybrids respectively.
Inclusion of a six-dimensional gluon condensate contribution does indeed seem to 
stabilize the sum-rule analysis.

The $Y(3940)$ (likely the same resonance as the $X(3915)$) seen by both the 
Belle~\cite{Choi:2005} and BaBar~\cite{Aubert:2007vj} collaborations is a
$C=+$ isosinglet with a mass of 3.915~GeV.  Unfortunately, at present, its
parity is unknown and its spin is uncertain: either $J=0$ or $J=2$.  As such,
the $Y(3940)$ \emph{could} be a pseudoscalar, but additional work is needed to
establish its $J^P$ assignment. 
In~\cite{Choi:2005}, this particle was already touted as a hybrid candidate based on both its
production mechanism and its observed decay mode.
The $Y(3940)$ is seen in $B$ decays which are thought to be particularly well-suited to hybrid 
production~\cite{Close:1997wp}.  Also, the $Y(3940)$ has not been seen to decay through kinematically
allowed $D\,\overline{D}^{(*)}$ modes, an observation consistent with a flux tube model-inspired
selection rule which suppresses hybrid decays to pairs of $S$-wave 
mesons~\cite{Page:1996rj,Close:1994hc,Isgur:1985vy}.  If the $Y(3940)$ is ultimately shown to have
$J^{PC}=0^{-+}$, then our mass prediction of $(3.82\pm0.13)$~GeV would provide additional
support in favour of a charmonium hybrid interpretation.

\bigskip
\noindent
{\bf Acknowledgements:}  We are grateful for financial support from the Natural Sciences and Engineering Research Council of Canada (NSERC).







\begin{thebibliography}{99}

\bibitem{Eidelman:2012vu}
  S.~Eidelman, B.~K.~Heltsley, J.~J.~Hernandez-Rey, S.~Navas and C.~Patrignani,
  arXiv:1205.4189 [hep-ex].

\bibitem{Barnes:1995hc}
  T.~Barnes, F.~E.~Close and E.~S.~Swanson,
  Phys.\ Rev.\ D {\bf 52} (1995) 5242
  [hep-ph/9501405].

  
\bibitem{Govaerts:1985fx} 
  J.~Govaerts, L.~J.~Reinders and J.~Weyers,
  Nucl.\ Phys.\ B {\bf 262} (1985) 575.

\bibitem{Govaerts:1984hc}
  J.~Govaerts, L.~J.~Reinders, H.~R.~Rubinstein and J.~Weyers,
  Nucl.\ Phys.\ B {\bf 258} (1985) 215.
  
\bibitem{Govaerts:1986pp}
  J.~Govaerts, L.~J.~Reinders, P.~Francken, X.~Gonze and J.~Weyers,
  Nucl.\ Phys.\ B {\bf 284} (1987) 674.
  
\bibitem{Qiao:2010zh}
  C.~-F.~Qiao, L.~Tang, G.~Hao and X.~-Q.~Li,
  J.\ Phys.\ G  {\bf 39} (2012) 015005
  [arXiv:1012.2614 [hep-ph]].
 
\bibitem{Berg:2012gd}
  R.~Berg, D.~Harnett, R.~T.~Kleiv and T.~G.~Steele,
  arXiv:1204.0049 [hep-ph].
  
\bibitem{Harnett:2012gs}
  D.~Harnett, R.~T.~Kleiv, T.~G.~Steele and H.~-y.~Jin,
  arXiv:1206.6776 [hep-ph].

\bibitem{Narison:2010cg}
  S.~Narison,
  Phys.\ Lett.\ B {\bf 693} (2010) 559
   [Erratum-ibid.\  {\bf 705} (2011) 544]
  [arXiv:1004.5333 [hep-ph]].
  
\bibitem{Binosi:2003yf}
  D.~Binosi and L.~Theussl,
  Comput.\ Phys.\ Commun.\  {\bf 161} (2004) 76
  [hep-ph/0309015].

\bibitem{Elias:1987ac}
  V.~Elias, T.~G.~Steele and M.~D.~Scadron,
  Phys.\ Rev.\ D {\bf 38} (1988) 1584. 
  
\bibitem{Bagan:1992tg}
  E.~Bagan, M.~R.~Ahmady, V.~Elias and T.~G.~Steele,
  Z.\ Phys.\ C {\bf 61} (1994) 157.

\bibitem{Mertig:1998vk}
  R.~Mertig and R.~Scharf,
  Comput.\ Phys.\ Commun.\  {\bf 111} (1998) 265
  [hep-ph/9801383].
  
\bibitem{Tarasov:1996br}
  O.~V.~Tarasov,
  Phys.\ Rev.\ D {\bf 54} (1996) 6479
  [hep-th/9606018].

\bibitem{Tarasov:1997kx}
  O.~V.~Tarasov,
  Nucl.\ Phys.\ B {\bf 502} (1997) 455
  [hep-ph/9703319].

\bibitem{Broadhurst:1993mw}
  D.~J.~Broadhurst, J.~Fleischer and O.~V.~Tarasov,
  Z.\ Phys.\ C {\bf 60} (1993) 287
  [hep-ph/9304303].
  
\bibitem{Davydychev:1991} 
  A.~I.~Davydychev,
  J.\ Math.\ Phys.\  {\bf 32}, 1052 (1991).

\bibitem{Bateman:1953}
A.~Erd\'{e}lyi,~(ed). 
``Higher Transcendental Functions'' (Bateman manuscript project), vol. 1. (McGraw-Hill, New York 1953).

\bibitem{Shifman:1978bx}
  M.~A.~Shifman, A.~I.~Vainshtein and V.~I.~Zakharov,
  Nucl.\ Phys.\ B {\bf 147} (1979) 385.

\bibitem{Shifman:1978by}
  M.~A.~Shifman, A.~I.~Vainshtein and V.~I.~Zakharov,
  Nucl.\ Phys.\ B {\bf 147} (1979) 448.
  
\bibitem{Bertlmann:1984ih}
  R.~A.~Bertlmann, G.~Launer and E.~de Rafael,
  Nucl.\ Phys.\ B {\bf 250} (1985) 61.

\bibitem{Harnett:2000xz}
  D.~Harnett, T.~G.~Steele and V.~Elias,
  Nucl.\ Phys.\ A {\bf 686} (2001) 393
  [hep-ph/0007049].
  
\bibitem{Chetyrkin:2009fv}
  K.~G.~Chetyrkin, J.~H.~Kuhn, A.~Maier, P.~Maierhofer, P.~Marquard, M.~Steinhauser and C.~Sturm,
  Phys.\ Rev.\ D {\bf 80} (2009) 074010
  [arXiv:0907.2110 [hep-ph]].
  
\bibitem{Narison:2011rn}
  S.~Narison,
  Phys.\ Lett.\ B {\bf 707} (2012) 259
  [arXiv:1105.5070 [hep-ph]].
  

\bibitem{Kuhn:2007vp}
  J.~H.~Kuhn, M.~Steinhauser and C.~Sturm,
  Nucl.\ Phys.\ B {\bf 778} (2007) 192
  [hep-ph/0702103].
  
\bibitem{pdg} Particle Data Group (K.~Nakamura et. al), J.~Phys.\ G {\bf 37} (2010) 075021.

\bibitem{Bethke:2009jm}
  S.~Bethke,
  Eur.\ Phys.\ J.\ C {\bf 64} (2009) 689
  [arXiv:0908.1135 [hep-ph]].
  
\bibitem{Narison:1981ts}
  S.~Narison and E.~de Rafael,
  Phys.\ Lett.\ B {\bf 103} (1981) 57.

\bibitem{Choi:2005} 
 S.-K.~Choi {\it et al.}  [Belle Collaboration]
Phys.\ Rev.\ Lett.\  {\bf 94} (2005) 182002
[arXiv:0408126 [hep-ex]].

\bibitem{Aubert:2007vj}
  B.~Aubert {\it et al.}  [BaBar Collaboration],
  Phys.\ Rev.\ Lett.\  {\bf 101} (2008) 082001
  [arXiv:0711.2047 [hep-ex]].


\bibitem{Close:1997wp}
  F.~E.~Close, I.~Dunietz, P.~R.~Page, S.~Veseli and H.~Yamamoto,
  Phys.\ Rev.\ D {\bf 57} (1998) 5653
  [hep-ph/9708265].


\bibitem{Page:1996rj}
  P.~R.~Page,
  Phys.\ Lett.\ B {\bf 402} (1997) 183
  [hep-ph/9611375].


\bibitem{Close:1994hc}
  F.~E.~Close and P.~R.~Page,
  Nucl.\ Phys.\ B {\bf 443} (1995) 233
  [hep-ph/9411301].

\bibitem{Isgur:1985vy}
  N.~Isgur, R.~Kokoski and J.~E.~Paton,
  Phys.\ Rev.\ Lett.\  {\bf 54} (1985) 869.
  
\end{thebibliography}







\end{document}